\documentstyle[epsf]{mn} 

\renewcommand{\d}{{\rm d}}
\newlength{\fsize}
\begin{document}

\title[QSO-galaxy correlations in arbitrary Friedmann cosmologies] 
      {QSO-galaxy correlations due to weak lensing in arbitrary
      Friedmann-Lema{\^\i}tre cosmologies}

\author[K.\ Dolag \& M.\ Bartelmann] 
       {Klaus Dolag and Matthias Bartelmann\\
	Max-Planck-Institut f\"ur Astrophysik, P.O.\ Box 1523,
        D--85740 Garching, Germany}

\maketitle

\begin{abstract}
We calculate the angular cross-correlation function between background
QSOs and foreground galaxies induced by the weak lensing effect of
large-scale structures. Results are given for arbitrary
Friedmann-Lema{\^\i}tre cosmologies. The non-linear growth of density
perturbations is included. Compared to the linear growth, the
non-linear growth increases the correlation amplitude by about an
order of magnitude in an Einstein-de Sitter universe, and by even more
for lower $\Omega_0$. The dependence of the correlation amplitude on
the cosmological parameters strongly depends on the normalization of
the power spectrum. The QSO-galaxy cross-correlation function is most
sensitive to density structures on scales in the range
$(1-10)\,h^{-1}\,$Mpc, where the normalization of the power spectrum
to the observed cluster abundance appears most appropriate. In that
case, the correlation strength changes by less than a factor of $\la2$
when $\Omega_0$ varies between $0.3$ and $1$, quite independent of the
value of $\Omega_\Lambda$. For $\Omega_0\la0.3$, the correlation
strength increases with decreasing $\Omega_0$, and it scales
approximately linearly with the Hubble constant $h$.
\end{abstract}

\section{Introduction}
\label{sec:1}

The existence of angular cross correlations between QSOs of moderate
or high redshift with luminous foreground material on angular scales
of $\ga10'$ has observationally been established in numerous
studies. Fugmann (1990) correlated optically bright, radio-loud QSOs
with Lick galaxies and found a significant overdensity of galaxies
around the QSOs of some of his QSO subsamples. Bartelmann \& Schneider
(1993b) repeated Fugmann's analysis with a well-defined sample of
background QSOs, confirming the correlation at the $98\%$ confidence
level for high-redshift, optically bright radio-QSOs. With a similar
correlation technique, correlations between the 1-Jansky QSO sample
and IRAS galaxies (Bartelmann \& Schneider 1994) and diffuse X-ray
emission (Bartelmann, Schneider, \& Hasinger 1994) were investigated,
yielding qualitatively the same results. Hutchings (1995) found
evidence for an excess at the $\ga5\,\sigma$ level of faint galaxies
in fields of $\sim1'$ radius around seven QSOs with $z=2.3$.

Rodrigues-Williams \& Hogan (1994) found a highly significant
correlation between optically selected, high-redshift QSOs and Zwicky
clusters. They discuss lensing as the most probable origin of the
correlations, although simple mass models for the clusters yield lower
magnifications than required to explain the significance of the
effect. Seitz \& Schneider (1995) repeated Rodrigues-Williams \&
Hogan's analysis with the 1-Jansky sample of QSOs. They found
agreement with the previous result for intermediate-redshift
($z\sim1$) QSOs, but failed to detect significant correlations for
higher-redshift sources. A variability-selected QSO sample was
correlated with Zwicky clusters by Rodrigues-Williams \& Hawkins
(1995). They detected a significant correlation between distant QSOs
and foreground Zwicky clusters, and interpreted it in terms of
gravitational lensing. Again, the implied average QSO magnification is
substantially larger than that inferred from simple lens models for
clusters with typical velocity dispersions. Wu \& Han (1995) searched
for associations between distant 1-Jansky and 2-Jansky QSOs and
foreground Abell clusters. They found no correlations with the
1-Jansky sources, and a marginally significant correlation with
2-Jansky sources. They argue that lensing by individual clusters is
insufficient for typical cluster velocity dispersions, and that
lensing by large-scale structures provides a viable explanation.

Ben{\'\i}tez \& Mart{\'\i}nez-Gonz\'alez (1995, 1997) found an excess
of red galaxies from the APM catalog with moderate-redshift 1-Jansky
QSOs on angular scales $<5'$ at the $99.1\%$ significance level. Their
colour selection ensures that the galaxies are most likely well in the
foreground of the QSOs. Bartsch, Schneider, \& Bartelmann (1997)
re-analyzed the correlation between IRAS galaxies and 1-Jansky QSOs
using a more advanced statistical technique which can be optimized to
the correlation function expected from lensing by large-scale
structures. In agreement with Bartelmann \& Schneider (1994), they
found significant correlations between the QSOs and the IRAS galaxies
on angular scales of $\sim5'$, but the correlation amplitude is higher
than expected from weak lensing by large-scale structures.

All these results indicate that there are correlations between
background QSOs and foreground ``light'', with ``light'' either in the
optical, the infrared, or the (soft) X-ray wavebands. Since the
foreground light emission is separated from the QSOs by typically a
few hundred Mpc's, physical associations are ruled out. Gravitational
lensing is then the most likely interpretation. By the magnification
bias, objects which are magnified by lensing are preferentially
included in flux-limited samples. A higher than average fraction of
these objects is therefore found in the background of matter
overdensities. If light traces the overdensities, a correlation
between foreground light and background QSOs can be established.

The angular scale of the correlations of a few arc minutes is
compatible with that expected from lensing by large-scale structures,
and the amplitude is either consistent with that explanation or
somewhat larger (Bartelmann \& Schneider 1993a). Wu \& Fang (1996)
discussed whether the autocorrelation of clusters modeled as singular
isothermal spheres can produce sufficient magnification to explain
this result. They found that this is not the case, and argue that
large-scale structures must contribute substantially. In any case, the
correlation scale is almost two orders of magnitude larger than the
typical Einstein radius of individual galaxies. Hence, correlations on
such scales cannot be attributed to the lensing effects of individual
galaxies alone.

If lensing is indeed responsible for the correlations detected, other
signatures of lensing should be found in the vicinity of such sources
which are correlated with foreground light. Indeed, Fort et al.\
(1996) searched for the shear expected from weak lensing in the fields
of five luminous QSOs and found coherent shear fields in all of
them. In addition, they detected galaxy groups in three of their
fields. Earlier, Bonnet et al.\ (1993) found evidence for coherent
weak shear in the field of the potentially multiply-imaged QSO
2345$+$007, which was later identified with a distant cluster (Mellier
et al.\ 1994; Fischer et al.\ 1994; Pell\'o et al.\ 1996).

Bartelmann (1995) has analytically calculated the angular
cross-correlation function between QSOs and galaxies assuming that (i)
lensing effects are weak and (ii) the biasing hypothesis of galaxy
formation (e.g.\ Kaiser 1984; Dekel \& Rees 1987) holds. This study
was restricted to (iii) linear growth of density perturbations and
(iv) Einstein-de Sitter (EdS) model universes. The goal of this paper
is to keep assumptions (i) and (ii), but give up the restrictions
(iii) and (iv). This will enable us to study the influence of
cosmological parameters on the correlation function, and to assess the
importance of non-linear growth of the density perturbations. Section
\ref{sec:2} reviews the formalism, part of which is derived in
appendix \ref{app:a}. We present the results in Sect.~\ref{sec:3} and
discuss them in Sect.~\ref{sec:4}.

\section{Formalism}
\label{sec:2}

As shown by Bartelmann (1995), the cross-correlation function
$\xi_{\rm QG}(\phi)$ between high-redshift QSOs and low-redshift
galaxies induced by weak gravitational lensing can be written
\begin{equation}
  \xi_{\rm QG}(\phi) = (s-1)\,b\,\xi_{\mu\delta}(\phi)\;.
\label{eq:1.1}
\end{equation}
As mentioned in the introduction, the assumptions underlying
eq.~(\ref{eq:1.1}) are that (i) lensing effects are weak (more
precisely, magnifications $|\mu-1|\ll1$), and that (ii) galaxies are
linearly biased relative to the dark matter by a factor $b$. $s$ is
the double-logarithmic slope of the differential QSO number counts
$n_{\rm Q}$ as a function of flux $S$, $n_{\rm Q}\propto S^{-s}$, and
$\xi_{\mu\delta}$ is the angular cross-correlation function between
magnification $\mu$ and density contrast $\delta$. The factor $(s-1)$
quantifies the magnification bias in eq.~(\ref{eq:1.1}). In order to
evaluate (\ref{eq:1.1}), we thus have to compute $\xi_{\mu\delta}$.

\subsection{Cross-correlation of magnification and density contrast}

A thin light bundle is sheared by the weak gravitational lensing
effect of the cosmic matter distribution by the ($2\times2$) tensor
$\gamma_{ij}$ derived in eq.~(\ref{eq:A.10}). Since lensing conserves
surface brightness, the light bundle's magnification is given by the
inverse determinant $\det^{-1}(\delta_{ij}+\gamma_{ij})$. Dealing with
weak lensing effects, we can assume $|\gamma_{ij}|\ll1$. Therefore,
the magnification $\mu=1+\delta\mu$ is
\begin{equation}
  1+\delta\mu = [\det(\delta_{ij}+\gamma_{ij})]^{-1} \approx 
  1 + {\rm tr}(\gamma_{ij})\;.
\label{eq:2.1}
\end{equation}

Via the trace of $\gamma_{ij}$, the magnification fluctuation contains
the trace of the ($2\times2$) Hessian $\Phi,_{ij}$ of the Newtonian
potential fluctuations $\Phi$. This can be augmented by $\Phi,_{33}$
because the potential derivative {\em along\/} the light ray does not
contribute to its deflection as long as the potential can be
considered static while it is being passed by the light ray. Thus,
$\nabla^2\Phi$ is introduced, which is related to the density contrast
$\delta$ through Poisson's equation. In comoving coordinates,
\begin{equation}
  \nabla^2\Phi = \frac{3\Omega_0}{2a}\,\delta\;,
\label{eq:2.2}
\end{equation}
where lengths are measured in units of the Hubble length $H_0^{-1}c$,
and the potential is scaled by $c^2$. With (\ref{eq:A.10}), the
magnification fluctuation $\delta\mu$ becomes
\begin{equation}
  \delta\mu(\vec w) = 3\Omega_0\,\int_0^w\,\d w'\,
  \frac{f_K(w')f_K(w-w')}{f_K(w)}\,\frac{\delta(\vec w')}{a}\;.
\label{eq:2.3}
\end{equation}
It depends on the (comoving) source distance $w$ and on the direction
$(\theta_1,\theta_2)$ into which the light ray starts off at the
observer. This is expressed in (\ref{eq:2.3}) by the vector $\vec
w=[f_K(w)\theta_1,f_K(w)\theta_2,w]$, with the curvature-dependent
radial distance $f_K(w)$ of (\ref{eq:A.8}).

The magnification fluctuation as a function of distance,
$\delta\mu(\vec w)$, has to be integrated along the line of sight,
weighted by the normalized distance distribution of the sources. Let
$W_{\rm Q}(w)$ be that distribution, then
\begin{equation}
  \delta\bar\mu(\theta_1,\theta_2) = 
  \int_0^\infty\,\d w'\,W_{\rm Q}(w')\,\delta\mu(\vec w')\;.
\label{eq:2.4}
\end{equation}
The upper bound on the integral is to be interpreted as $w$ at
$z=\infty$, which can be finite for some combinations of cosmological
parameters. By changing the order of integrations, (\ref{eq:2.4}) can
be written
\begin{equation}
  \delta\bar\mu(\theta_1,\theta_2) = 3\,\Omega_0\,
  \int_0^\infty\,\d w\,G_{\rm Q}(w)\,\frac{\delta(\vec w)}{a}
\label{eq:2.5}
\end{equation}
with the modified weight function
\begin{equation}
  G_{\rm Q}(w) = f_K(w)\,\int_w^\infty\,\d w'\,
  \frac{f_K(w'-w)}{f_K(w')}\,W_{\rm Q}(w')\;.
\label{eq:2.6}
\end{equation}

Likewise, the density contrast must be integrated along the
line-of-sight, weighted by the distance distribution of the galaxies
$W_{\rm G}(w)$ in order to get the projected galaxy number density on
the sky, according to the biasing hypothesis. Hence,
\begin{equation}
  \bar\delta(\theta_1,\theta_2) = 
  \int_0^\infty\,\d w\,\delta(\vec w)\,W_{\rm G}(w)\;.
\label{eq:2.7}
\end{equation}
For the shape of the weight functions $W_{\rm G}(w)$ and $G_{\rm
Q}(w)$, see Fig.~\ref{fig:1}.

Both $\bar\delta$ and $\delta\bar\mu$ are projections of the
three-dimensional density contrast. The density contrast is supposed
to be a homogeneous and isotropic random field characterized by a
power spectrum $P_\delta(k)$. Typical scales in $P_\delta(k)$ are on
the order of several ten Mpc's. The projections along the
line-of-sight have length scales comparable to the Hubble
radius. Under these circumstances, the Fourier-space analog of
Limber's equation can be applied (Kaiser 1992). It asserts that the
power spectrum $P_p(\kappa)$ of a projection $p$ of a
three-dimensional field $f$ with weight function $q$ is related to the
power spectrum $P_f(k)$ of $f$ through
\begin{equation}
  P_p(\kappa) = \int_0^\infty\,\d w\,\frac{q^2(w)}{f_K^2(w)}\,
  P_f\left(\frac{\kappa}{f_K(w)}\right)\;.
\label{eq:2.8}
\end{equation}
Note that $\vec\kappa$ is a two-dimensional wave vector, while $\vec
k$ is three-dimensional. The cross power spectrum
$P_{\mu\delta}(\kappa)$ of the density contrast and the magnification
fluctuation is therefore given by
\begin{equation}
  P_{\mu\delta}(\kappa) = 3\,\Omega_0\,
  \int_0^\infty\,\d w\,\frac{G_{\rm Q}(w)W_{\rm G}(w)}
  {a(w)\,f_K^2(w)}\,P_\delta\left(\frac{\kappa}{f_K(w)}\right)\;.
\label{eq:2.9}
\end{equation}

The cross correlation function $\xi_{\mu\delta}(\phi)$ is the Fourier
transform of $P_{\mu\delta}(\kappa)$,
\begin{equation}
  \xi_{\mu\delta}(\phi) =
  \int_0^\infty\frac{\kappa\d\kappa}{(2\pi)^2}\,
  P_{\mu\delta}(\kappa)\,\int_0^{2\pi}d\vartheta\,
  \exp(-{\rm i}\,\kappa\phi\cos\vartheta)\;,
\label{eq:2.10}
\end{equation}
where $\vartheta$ is the angle enclosed by $\vec\kappa$ and
$\vec\phi$. The $\vartheta$-integration yields a zeroth-order Bessel
function ${\rm J}_0$, and inserting (\ref{eq:2.9}) leads to
\begin{eqnarray}
  \xi_{\mu\delta}(\phi) &=& 3\,\Omega_0\,
  \int_0^\infty\,\d w\,\frac{G_{\rm Q}(w)W_{\rm G}(w)}{a(w)}
  \nonumber\\ &\times&
  \int_0^\infty\,\frac{k\d k}{2\pi}\,P_\delta(k,w)\,
  {\rm J}_0[f_K(w)k\phi]\;.
\label{eq:2.11}
\end{eqnarray}
It should be noted that eq.~(\ref{eq:2.11}) automatically accounts for
two effects: First, matter inhomogeneities ``behind'' the QSOs do not
contribute to the gravitational lensing effects. This can be seen by
inserting $W_{\rm Q}=\delta(w-w_{\rm Q})$ into eq.~(\ref{eq:2.6}),
upon which $G_{\rm Q}\propto{\rm H}(w_{\rm Q}-w)$, with the Heaviside
step function ${\rm H}(x)$. Inserting this together with $W_{\rm
G}=\delta(w-w_{\rm G})$ with $w_{\rm G}>w_{\rm Q}$ into
eq.~(\ref{eq:2.11}) yields $\xi_{\mu\delta}=0$, as it should. Second,
the autocorrelation of galaxies is included through the matter power
spectrum $P_\delta(k)$.

The matter power spectrum $P_\delta(k,w)$ depends on time, which is
expressed in (\ref{eq:2.11}) as a dependence on comoving distance
$w$. Assuming linear growth of the density fluctuations,
$P_\delta(k,w)=P_\delta^0(k)\,D(w)$, with the power spectrum
$P_\delta^0(k)$ at the present epoch and a cosmology-dependent growth
function $D(w)$. An accurate fit formula for $D(w)$ is given in
Carroll, Press, \& Turner (1992). In that case, the integrations over
$w$ and $k$ can be separated. This yields the instructive equation
\begin{equation}
  \xi_{\mu\delta}(\phi) =
  \int_0^\infty\,\frac{k\d k}{2\pi}\,P_\delta(k,w)\,F(k,\phi)\;,
\label{eq:2.12}
\end{equation}
which shows that the cross correlation function between magnification
and density contrast is a convolution of the matter power spectrum
with a filter function $F(k,\phi)$, weighted by the wave number
$k$. For the Einstein-de Sitter case, $F(k,\phi)$ is plotted in
Bartelmann (1995; Fig.~2). For small angles $\phi$, $F(k,\phi)$
filters out scales smaller than $\sim1\,$Mpc. Here, we evaluate
eq.~(\ref{eq:2.11}) numerically, including the non-linear growth of
$P_\delta(k,w)$.

\subsection{Choice of the weight functions}

\begin{figure} 
\begin{center}
  \mbox{\epsfxsize=\fsize\epsffile{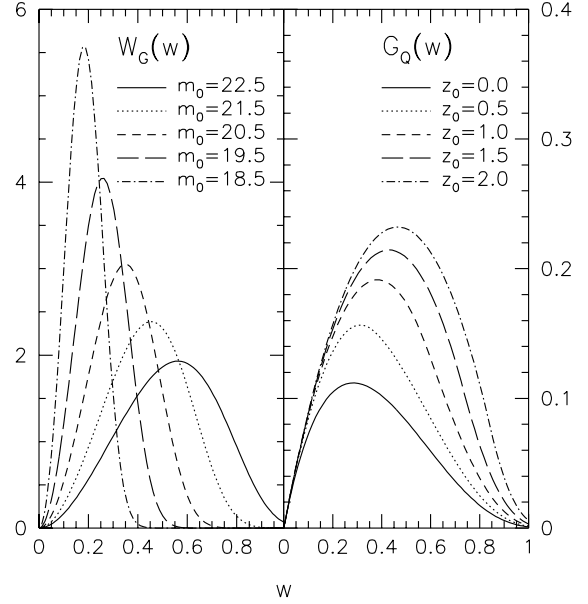}}
\end{center}
\caption{Examples for the galaxy weight function $W_{\rm G}(w)$ (left
  panel), and for the QSO weight functions $G_{\rm Q}(w)$ adopted from
  Pei (1994; right panel). All curves are shown for an Einstein-de
  Sitter model universe. $m_0$ is the detection threshold magnitude
  for the galaxies, and $z_0$ is the lower redshift cutoff of the QSO
  sample. Line types distinguish results for different $m_0$ or $z_0$,
  as indicated.}
\label{fig:1}
\end{figure}

We have to specify the QSO- and galaxy weight functions $G_{\rm Q}$
and $W_{\rm G}$. For the QSO weight function $W_{\rm Q}$, we choose a
QSO redshift distribution of the form given by Pei (1994), with
parameters adapted to observations in the same paper. With
(\ref{eq:2.6}), this yields QSO weight functions $G_{\rm Q}(w)$ for
which examples are given in the right panel of
Fig.~\ref{fig:1}. $G_{\rm Q}$ depends on the lower redshift cutoff
$z_0$ imposed on the QSO sample.

For the galaxy redshift distribution underlying the galaxy weight
function $W_{\rm G}(w)$, we assume a constant comoving number density
of galaxies. The galaxies are distributed in luminosity according to a
Schechter luminosity function with parameters $\nu=-0.81$ and
$L_*=1.1\times10^{10}L_\odot h^{-2}$. We further choose a threshold
apparent magnitude $m_0$ for the galaxies, which we calculate from $L$
and $w$ using a $K$-correction derived from an effective spectral
energy distribution with index $\alpha=-1.5$. Examples for the
resulting weight functions are given in the left panel of
Fig.~\ref{fig:1}. Our results for the cross-correlation function
$\xi_{\mu\delta}(\phi)$ show that $\xi_{\mu\delta}(\phi)$ depends only
very weakly on the exact shape of the weight functions $W_{\rm G}(w)$
and $G_{\rm Q}(w)$, once $m_0$ and $z_0$ are fixed.

\subsection{Matter power spectra}

We use in the following three different types of matter power spectra
$P_\delta(k)$, viz.\ cold dark matter (CDM), hot dark matter (HDM),
and mixed dark matter (MDM). For each of these dark-matter types,
several fitting formulae have been given in the literature. Comparing
results for the QSO-galaxy cross-correlation function, we found that
different formulae for the same dark-matter type yield only marginally
different correlation functions. We can therefore restrict the study
to one formula for each dark-matter type only. For definiteness, we
use those given by Efstathiou, Bond, \& White (1992) for CDM, Bardeen
et al.\ (1986) for HDM, and Holtzman (1989) for MDM.

\begin{figure} 
\begin{center}
  \mbox{\epsfxsize=\fsize\epsffile{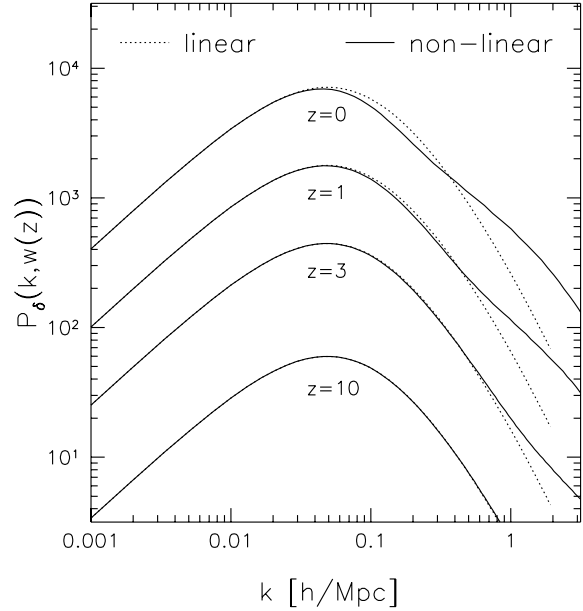}}
\end{center}
\caption{Comparison between the linear (dotted curves) and non-linear
  (solid curves) evolution of a CDM power spectrum, according to the
  prescription given by Peacock \& Dodds (1996) for an Einstein-de
  Sitter model universe. Non-linear power builds up on small scales at
  the expense of power on intermediate scales. The linear spectra are
  normalized to $\sigma_8=1$.}
\label{fig:2}
\end{figure}

Semi-analytic prescriptions for the non-linear evolution of the power
spectra have been given by Jain, Mo, \& White (1996) and Peacock \&
Dodds (1996). Both rest on the conjecture by Hamilton et al. (1991)
that the {\em non-linear\/} power on a given scale $k_{\rm nl}$ is
identified with the {\em linear\/} power on a larger scale $k_{\rm l}$
which is some universal function of $k_{\rm nl}$. This universal
function has to be identified through numerical simulations. The
prescription by Peacock \& Dodds (1996) works as long as the local
double-logarithmic slope of the power spectrum at the scale of
interest is not steeper than that of the CDM spectrum. This is not the
case at small scales for an HDM power spectrum. For that, we rather
use the evolution calculated in the framework of the Zel'dovich
approximation by Schneider \& Bartelmann (1995). For MDM, we use
linear evolution only. To give an example, Fig.~\ref{fig:2} shows the
non-linear evolution of a CDM spectrum in an Einstein-de Sitter
universe.

One further point in specifying the power spectra is to normalize the
shape predicted for the chosen matter model with measurements. There
are three basic possibilities to do this. (i) Determine the variance
of galaxy counts on a fixed scale, commonly $8\,h^{-1}\,$Mpc, and
assume that light follows mass, with some linear bias factor $b$. (ii)
Measure the abundance of galaxy clusters and fix the amplitude of the
power spectrum such that this abundance is reproduced. (iii) Fix the
amplitude of the power spectrum such that the amplitude of the cosmic
microwave background anisotropies are reproduced. It is well known
that these methods yield different amplitudes for the CDM power
spectrum. It is possible to match (i) and (ii) by choosing the bias
factor suitably. To also match (i) or (ii) with (iii), the shape of
the power spectra needs to be changed on small scales. We emphasize
here that the angular cross-correlation function between QSOs and
galaxies provides in principle a more direct measurement of the matter
power spectrum than counting galaxies or clusters, for it is based on
gravitational lensing. We shall later demonstrate the huge effect
which the non-linear growth of density fluctuations has on the
amplitude of the cross-correlation function on small angular scales
(Fig.~\ref{fig:2}).

We show results for all three normalizations of the power spectrum,
not accounting for the further constraints that are derived from
galaxy counts, galaxy autocorrelation functions, or the cluster
abundance. In that way, our results can easily be scaled to the
desired normalization.

\section{Results}
\label{sec:3}

As shown by Bartelmann (1995) for the special case of a linearly
evolving density field in an Einstein-de Sitter universe, the shape of
the cross-correlation function $\xi_{\mu\delta}(\phi)$ depends
sensitively on the type of the dark-matter particles. This property of
$\xi_{\mu\delta}(\phi)$ continues to hold for all the generalizations
we calculate here. We use the results obtained for linear density
evolution in an Einstein-de Sitter universe as a reference for the
influence of cosmological parameters and non-linear density evolution.

\subsection{Dependence on type and evolution of the dark matter}

\begin{figure} 
\begin{center}
  \mbox{\epsfxsize=\fsize\epsffile{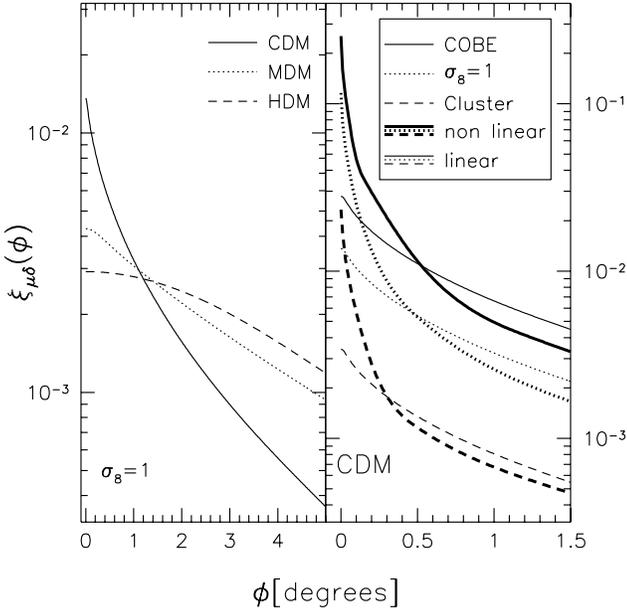}}
\end{center}
\caption{The cross-correlation function $\xi_{\mu\delta}(\phi)$ is
  shown here for an Einstein-de Sitter universe. Left panel: results
  for different types of {\em linearly\/} evolving dark-matter power
  spectra (CDM, solid curve; MDM, dotted curve; HDM, dashed
  curve). All spectra were normalized to $\sigma_8=1$. While the
  cross-correlation function peaks towards $\phi\to0$ for CDM, it
  flattens for HDM and MDM due to the cut-off at large wavenumbers $k$
  in those spectra. At large angles $\phi$, the HDM and MDM
  correlation functions lie above the CDM results. Right panel: the
  effect of the {\em non-linear\/} evolution of the CDM power spectrum
  on $\xi_{\mu\delta}(\phi)$, for three different normalizations, as
  indicated by line type. The COBE normalization yields the largest
  amplitude, the cluster normalization the smallest. The linear
  results are shown as the thin lines for comparison. For this choice
  of the cosmological model, non-linear evolution increases the
  correlation amplitude $\xi_{\mu\delta}(0)$ by about an order of
  magnitude, quite independent of the normalization. $m_0=20.5$ and
  $z_0=0.3$ were chosen.}
\label{fig:3}
\end{figure}

Figure \ref{fig:3} demonstrates the different shapes of
$\xi_{\mu\delta}(\phi)$ for different matter power spectra. In
agreement with the analytic description of Bartelmann (1995), the
cross-correlation function (left panel of Fig.~\ref{fig:3}) peaks for
$\phi\to0$, whereas it flattens off at small $\phi$ for
HDM. Power-series expansion yields $\xi_{\mu\delta}\propto\phi+{\cal
O}(\phi^2)$ for CDM, and $\xi_{\mu\delta}\propto\phi^2+{\cal
O}(\phi^4)$ for HDM (Bartelmann 1995). The reason for this is the
different asymptotic behavior of $P_\delta(k)$ for $k\to\infty$: the
cut-off in the HDM power spectrum is responsible for the flattening of
the HDM cross-correlation function. As expected, $\xi_{\mu\delta}$ for
MDM falls between these two cases, in that it flattens at smaller
$\phi$ than for HDM.

The cross-correlation function for the different normalizations of the
CDM power spectrum is shown in the right panel of
Fig.~\ref{fig:3}. Clearly, the high normalization to the COBE results
yields the strongest correlation, while the normalization to cluster
counts yields the weakest. For small angles $\phi$, the window
function $F(k,\phi)$ (see Bartelmann 1995 for details) drops to zero
at the scale of galaxy clusters and smaller structures. Hence, the
integral (\ref{eq:2.11}) yields the strongest correlation for the COBE
normalization, because the COBE normalization leads to much more power
on cluster scales than the cluster normalization.

The non-linear growth of the density fluctuations also most strongly
affects the power on scales of clusters and smaller (see
Fig.~\ref{fig:2}). Because of that, the influence of non-linear
density evolution on the cross-correlation function is huge. The right
panel of Fig.~\ref{fig:3} contrasts the cross-correlation function for
non-linear and linear density evolution of the power spectra. For the
CDM model in an Einstein-de Sitter universe, the amplitude,
$\xi_{\mu\delta}(0)$, of the cross-correlation function increases by
about one order in magnitude, quite independent of the
normalization. Depending on the normalization, the increase can be
even larger in other model universes. Since $\xi_{\mu\delta}(\phi)$
remains unaffected on large angular scales, the peak for $\phi\to0$ is
sharpened by non-linear evolution, in tentative agreement with
observations (Ben{\'\i}tez \& Mart{\'\i}nez-Gonz\'alez 1995;
1997). Non-linear evolution of the power spectra is, therefore,
crucial to match theoretical expectations and measurements.

Non-linear evolution also increases the HDM correlation
amplitude. However, the flattening of the HDM correlation function for
$\phi\to0$, and its much lower amplitude than for CDM, appears to be
ruled out by observations, and we will therefore concentrate on CDM
spectra in the following.

\subsection{Dependence on cosmological parameters}

The change of the cross-correlation function with different
cosmological parameters shows a huge dependence on the normalization
of the power spectrum. This is due to the fact that the three
normalizations themselves depend differently on the cosmological
parameters.

\begin{figure} 
\begin{center}
  \mbox{\epsfxsize=\fsize\epsffile{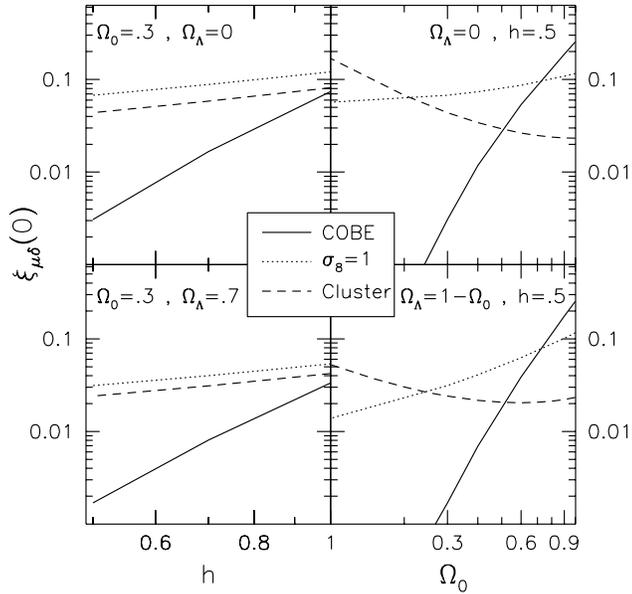}}
\end{center}
\caption{The dependence of the correlation amplitude
  $\xi_{\mu\delta}(0)$ on the Hubble constant $h$ and the matter
  density $\Omega_0$ is shown for different cosmological models and
  different normalizations of the CDM power spectrum. Non-linear
  evolution was assumed. Left panels: dependence on $h$; right panels:
  dependence on $\Omega_0$; upper panels: open models
  ($\Omega_\Lambda=0$); lower panels: spatially flat models
  ($\Omega_0+\Omega_\Lambda=1$). Different normalizations are
  distinguished by line type, as indicated. The COBE normalization
  yields the strongest dependence on both $h$ and
  $\Omega_0$. Generally, $\xi_{\mu\delta}(0)$ decreases with
  decreasing $h$ and decreasing $\Omega_0$, but note that the cluster
  normalization reverses that trend, especially for the open models
  (upper right panel). All curves were calculated assuming $m_0=20.5$
  and $z_0=0.3$.}
\label{fig:4}
\end{figure}

To understand the dependence of the correlation function on the
cosmological parameters a bit better, we adapt an analytical
expression by Bartelmann (1995) to our discussion. Using a CDM model
spectrum, and assuming linear density evolution, Bartelmann (1995)
obtains
\begin{equation}
  \xi_{\mu\delta}(0)=\frac{3\,A_{\rm CDM}\,k_0^3}{2}\,K_0
  \label{eq:3.1}
\end{equation} 
for an Einstein-de Sitter universe. Here, $A_{\rm CDM}$ is the
amplitude of the power spectrum, $k_0$ is the wave number of the peak
in the spectrum, and $K_0$ is an integral over the weight functions
introduced in Sect.~2.2. In non-EdS cosmologies, we get an additional
factor $\Omega_0$ from Poisson's equation, and $k_0\propto\Omega_0\,h$
in units of the inverse Hubble length. Using this, the dependence of
the amplitude $\xi_{\mu\delta}(0)$ (for linear evolution) can be
written as
\begin{equation}
  \xi_{\mu\delta}(0)\propto\Omega_0^4\,h^3\times
  [A_{\rm CDM}]\times[K_0]\;,
\label{eq:3.2}
\end{equation}
where the square brackets around a symbol denote the dependence of the
corresponding quantity on the cosmological
parameters. Fig.~\ref{fig:4} illustrates the dependence of
$\xi_{\mu\delta}(0)$ on $h$ and $\Omega_0$ for various cosmological
models and normalizations of the power spectrum.

\begin{figure} 
\begin{center}
  \mbox{\epsfxsize=\fsize\epsffile{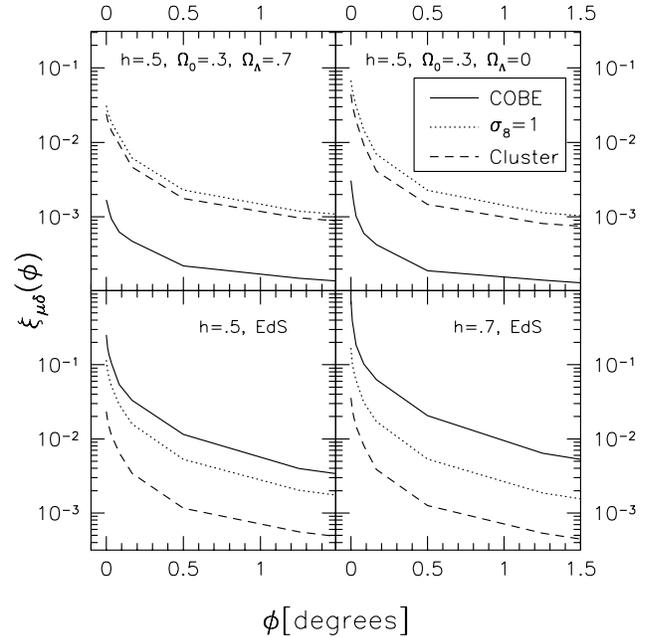}}
\end{center}
\caption{The angular cross-correlation function
  $\xi_{\mu\delta}(\phi)$ is plotted for different values of the
  cosmological parameters. In all panels, curves for the three
  normalizations are shown, distinguished by line type as
  indicated. Top left panel: spatially flat low-density universe with
  $h=0.5$, $\Omega_0=0.3$, and $\Omega_\Lambda=0.7$; top right panel:
  open low-density universe with $h=0.5$, $\Omega_0=0.3$, and
  $\Omega_\Lambda=0$; bottom panels: Einstein-de Sitter universes with
  $h=0.5$ (left) and $h=0.7$ (right). All curves are for CDM and for
  $m_0=20.5$ and $z_0=0.3$.}
\label{fig:5}
\end{figure}

We looked in detail at three different cosmological models. These are:
an Einstein-de Sitter universe with two different values of the Hubble
constant $H_0$, an open low-density model with $\Omega_0=0.3$, and
$\Omega_\Lambda=0$, and a spatially flat low-density model with
$\Omega_0=0.3$ and $\Omega_\Lambda=0.7$. The correlation functions
$\xi_{\mu\delta}(\phi)$ for these cosmologies and different
normalizations are shown in Fig.~\ref{fig:5}. The left panel of
Fig.~\ref{fig:6} illustrates the dependence on $\Omega_0$ of the
correlation amplitude $\xi_{\rm QG}(0)$ for cluster normalization. We
restrict the discussion of the dependence on cosmological parameters
to non-linear models only.

In an Einstein-de Sitter universe, the correlation is strongest for
the COBE normalization. This changes in low-density universes, where
we find a stronger correlation for the cluster- and the $\sigma_8=1$
normalizations. Generally, the correlation amplitude increases with
$\Omega_0$, except for the cluster normalization. There, the
correlation amplitude changes only mildly with $\Omega_0$ as long as
$\Omega_0\ga0.3$, and it {\em increases\/} when $\Omega_0$ decreases
even further. Likewise, a larger Hubble constant yields a higher
correlation amplitude. Both these effects are mainly due to the fact
that the peak in the power spectrum shifts towards larger wave numbers
with increasing $\Omega_0$ and $H_0$, and the additional factor
$\Omega_0$ from Poisson's equation renders a stronger dependence on
$\Omega_0$ than on $H_0$. Depending itself quite strongly on
$\Omega_0$, the cluster normalization can reverse the general trend
with $\Omega_0$.

Approximate relations between the amplitude $\xi_{\mu\delta}(0)$ and
$\Omega_0$ and $H_0$ are shown in Tab.~\ref{tab:1}. They are in good
qualitative agreement with the expectations from
eq.~(\ref{eq:3.2}). Figure \ref{fig:4} and Tab.~\ref{tab:1} also show
that the strongest dependence on the cosmological parameters is
obtained for the COBE normalization. The dependence on the
cosmological constant $\Omega_\Lambda$ for fixed $\Omega_0$ is much
weaker than for the other parameters, which reflects the fact that
$\Omega_\Lambda$ affects only the space-time curvature, which enters
through $K_0$ in eq.~(\ref{eq:3.2}).

\begin{table}
\caption{This table summarizes the dependence of the correlation
  amplitude $\xi_{\mu\delta}(0)$ on the cosmological parameters $h$
  and $\Omega_0$, for non-linearly evolved CDM spectra. We fitted
  parabolae in $\ln\xi_{\mu\delta}(0)$--$\ln Q$ space to the curves in
  Fig.~\protect\ref{fig:4}, with $Q$ either $h$ or $\Omega_0$. Hence,
  we fit $\ln\xi_{\mu\delta}(0)=A\,(\ln Q)^2+B\,\ln Q+C$, and give the
  coefficients $A$, $B$ and $C$ in the table. When $h$ is varied, we
  take $\Omega_0=0.3$, $\Omega_\Lambda=0$ for the open models, and
  $\Omega_0=0.3$, $\Omega_\Lambda=0.7$ for the spatially flat
  models. When $\Omega_0$ is varied, we take $h=0.5$. The dependences
  differ with the normalization, hence we give results for the three
  normalizations we have chosen. For the cluster normalization, we
  also give results for the correlation amplitude $\xi_{\rm QG}(0)$,
  which differs from $\xi_{\mu\delta}(0)$ by the bias factor
  $b=\sigma_8^{-1}$.}
\label{tab:1}
\begin{center}
\begin{tabular}{llrrrr} 
\hline
 & & 
\multicolumn{4}{c}{Normalization} \\ 
\cline{3-6}
$Q$ & Coeff. & COBE & $\sigma_8=1$ &
\multicolumn{2}{c}{Cluster} \\ 
\cline{5-6}
 & & $\xi_{\mu\delta}(0)$ &  $\xi_{\mu\delta}(0)$ &
 $\xi_{\mu\delta}(0)$ & $\xi_{\rm QG}(0)$ \\
\hline
           & $A$ & $-1.2$ & $ 0.1$ & $ 0.1$ & $ 0.1$ \\
$h$        & $B$ & $ 3.7$ & $ 0.9$ & $ 1.0$ & $ 1.0$ \\
(open)     & $C$ & $-2.5$ & $-2.1$ & $-2.5$ & $-1.8$ \\
\hline	                                             
           & $A$ & $-1.0$ & $ 0.1$ & $ 0.1$ & $ 0.1$ \\
$h$        & $B$ & $ 3.6$ & $ 0.9$ & $ 0.9$ & $ 0.9$ \\
(flat)     & $C$ & $-3.5$ & $-3.0$ & $-3.2$ & $-2.5$ \\
\hline	                                             
           & $A$ & $-1.0$ & $ 0.1$ & $ 0.3$ & $ 0.3$ \\
$\Omega_0$ & $B$ & $ 2.5$ & $ 0.6$ & $-0.2$ & $ 0.3$ \\
(open)     & $C$ & $-1.4$ & $-2.1$ & $-3.7$ & $-2.5$ \\
\hline	                                             
           & $A$ & $-0.3$ & $ 0.2$ & $ 0.3$ & $ 0.3$ \\
$\Omega_0$ & $B$ & $ 3.8$ & $ 1.3$ & $ 0.3$ & $ 0.8$ \\
(flat)     & $C$ & $-1.4$ & $-2.1$ & $-3.7$ & $-2.5$ \\
\hline
\end{tabular}
\end{center}
\end{table}

\subsection{Other model parameters}

\begin{figure} 
\begin{center}
  \mbox{\epsfxsize=\fsize\epsffile{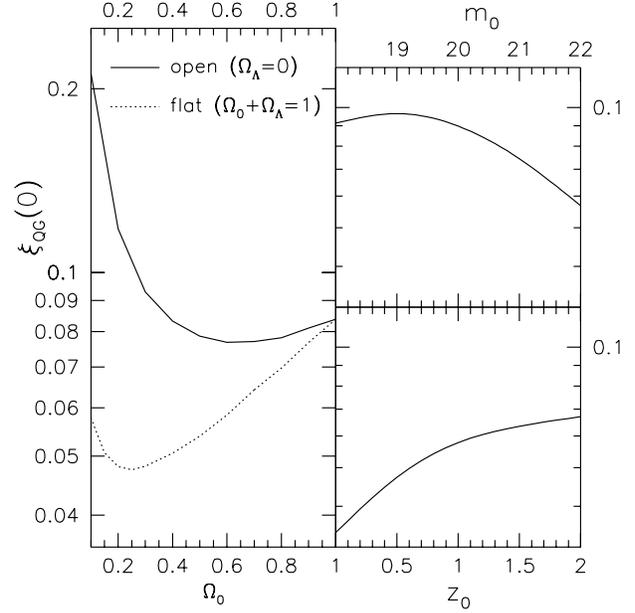}}
\end{center}
\caption{Left panel: amplitude of the angular cross-correlation
  function between QSOs and galaxies, $\xi_{\rm QG}(0)$, as a function
  of $\Omega_0$. The curves are for cluster-normalized, non-linearly
  evolving CDM in open and spatially flat model universes. Note the
  difference to the correlation amplitude $\xi_{\mu\delta}(0)$ between
  magnification and density contrast: We have assumed a bias factor of
  $b=\sigma_8^{-1}$ here. Upper right panel: dependence of $\xi_{\rm
  QG}(0)$ on the threshold galaxy magnitude $m_0$; lower right panel:
  dependence of $\xi_{\rm QG}(0)$ on the minimum QSO redshift
  $z_0$. While $\xi_{\rm QG}(0)$ increases monotonically with $z_0$,
  it reaches a maximum for $m_0\simeq19$ and decreases for fainter
  galaxies. The curves in the right panels are calculated in an
  Einstein-de Sitter universe with cluster normalization of the power
  spectrum. $z_0=0.3$ in the upper right panel and $m_0=20.5$ in the
  lower right panel.}
\label{fig:6}
\end{figure}

Quite generally, the dependence of the correlation function on all the
other parameters entering our calculation is weak. As mentioned in
Sect.~2.3, there is no significant change of the correlation function
upon using different spectra for the same dark-matter model. Small
changes in the amplitude of the angular cross-correlation function
result from varying the parameters $z_0$ and $m_0$ of the weight
functions introduced in Sect.~2.2. This is due to the fact that
varying the mean redshift of the galaxies (through $m_0$), or the QSO
redshift distribution (through $z_0$) changes the lensing
efficiency. Examples are shown in the right panels of
Fig.~\ref{fig:6}. The upper right panel illustrates how the
correlation amplitude changes if the mean redshift of the galaxies is
shifted beyond the maximum in lensing efficiency for a fixed QSO
redshift distribution. The lower right panel shows how the correlation
amplitude increases with increasing QSO redshift.

The fact that $\xi_{\mu\delta}(\phi)$ peaks for moderately faint
galaxies rather than for the faintest can easily be understood. The
galaxies are simply tracers of the density inhomogeneities that act as
lenses on the QSO samples. At fixed QSO redshift, there is an
intermediate redshift at which the lenses are most efficient. This
redshift is somewhat reduced by the redshift dependence of structure
growth. Allowing for fainter galaxies selects more distant density
inhomogeneities which are less efficient lenses, making the
cross-correlation function $\xi_{\mu\delta}$ drop.

\section{Summary and discussion}
\label{sec:4}

We have extended an earlier study of the angular cross-correlation
function $\xi_{\rm QG}(\phi)$ between QSOs and galaxies (Bartelmann
1995) in three respects. First, we generalize from the Einstein-de
Sitter to arbitrary Friedmann-Lema{\^\i}tre cosmological models,
second, we allow for the non-linear evolution of the density
fluctuations, and third, we construct more realistic redshift
distributions for galaxies and QSOs. The QSO-galaxy correlation
function is expressed in terms of the correlation function between
magnification and density contrast, $\xi_{\mu\delta}(\phi)$.

We confirm that the shape of the cross-correlation function is very
sensitive to the type of the dark-matter power spectrum: While it
peaks for $\phi\to0$ in the case of CDM, it flattens for HDM and
MDM. Measurements of $\xi_{\rm QG}(\phi)$ indicate that the
correlation peaks for $\phi\to0$, in contradiction to the flattening
expected for HDM. Given further the much lower correlation amplitude
for HDM compared to CDM, we focus on the CDM model.

We summarize our main results as follows:

\begin{enumerate}

\item Non-linear evolution of the density fluctuations leads to an
increase in the correlation amplitude $\xi_{\mu\delta}(0)$ relative to
the linear case by about an order of magnitude, or more.

\item $\xi_{\mu\delta}(\phi)$ depends on the cosmological parameters
$\Omega_0$, $H_0$, and $\Omega_\Lambda$ in different ways:

\begin{itemize}

\item $\Omega_0$ enters through Poisson's equation, through the
normalization, the shape, and the evolution of the power spectrum, and
through space-time curvature. The location of the peak in the power
spectrum is proportional to $\Omega_0\,h=\Gamma$. The larger $\Gamma$
is, the larger is the overlap between the power spectrum and the
window function that filters out the scales relevant for
$\xi_{\mu\delta}$. Therefore, the correlation generally becomes
stronger with increasing $\Omega_0$, except for the cluster
normalization.

\item The Hubble constant enters through the shape parameter of the
power spectrum, $\Gamma=\Omega_0\,h$. As for $\Omega_0$, the
correlation strengthens with increasing $H_0$.

\item The cosmological constant $\Omega_\Lambda$ enters only weakly
through space-time curvature and the evolution of the power spectrum.

\end{itemize}

\item The method by which the power spectrum is normalized plays a
critical role, both for the amplitude of the correlation and its
dependence on cosmological parameters. When the spectrum is normalized
to the COBE measurements, the dependence on cosmological parameters is
strongest. For the cluster normalization, the dependence on $\Omega_0$
is weakest, and it can even be reversed for small $\Omega_0$, because
the general decrease is compensated by the $\Omega_0$ dependence of
the normalization.

\item Other parameters entering the calculation of the correlation
function have a comparably modest influence. Factors of $1.5-2$ can be
gained or lost by varying the galaxy detection threshold $m_0$ or the
minimum QSO redshift $z_0$. The exact shape of the galaxy- and QSO
redshift distributions is largely irrelevant, and different
representations published in the literature for the same type of power
spectrum yield the same correlation functions.

\end{enumerate}

\begin{figure} 
\begin{center}
  \mbox{\epsfxsize=\fsize\epsffile{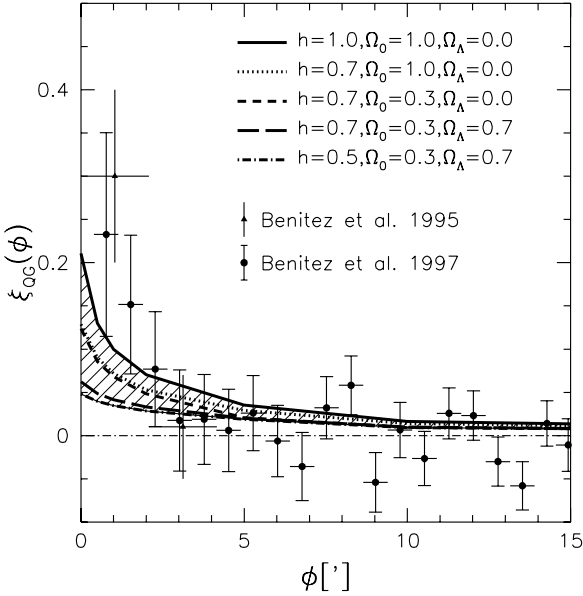}}
\end{center}
\caption{The range of correlation functions $\xi_{\rm QG}(\phi)$ that
  can be covered by varying cosmological parameters while keeping the
  normalization of the power spectrum fixed to the cluster
  abundance. As Tab.~\protect\ref{tab:1} and Fig.~\protect\ref{fig:4}
  show for that normalization, $\xi_{\mu\delta}(\phi)$ changes by less
  than a factor $\protect\la2$ for $\Omega_0\protect\ga0.3$, and it
  scales with $H_0$ approximately linearly. Note that we are plotting
  $\xi_{\rm QG}$ rather than $\xi_{\mu\delta}$ here. We have assumed
  $s=2.8$ (cf.\ Bartelmann 1995) and $b=\sigma_8^{-1}$ to relate the
  two; see eq.~(\protect\ref{eq:1.1}). The cluster normalization
  requires roughly $\sigma_8\approx0.6\,\Omega_0^{0.6}$, hence
  $b\propto\Omega_0^{-0.6}$. See also the curves in the left panel of
  Fig.~\protect\ref{fig:6}. We have further assumed $m_0=20.5$ and
  $z_0=0.3$, corresponding to the values chosen in Ben{\'\i}tez \&
  Mart{\'\i}nez-Gonz\'alez (1995, 1997), whose data points are added
  for illustration.}
\label{fig:7}
\end{figure}

We have argued that, for small angular scales $\phi$,
$\xi_{\mu\delta}(\phi)$ is most sensitive to the power on cluster
scales. This is because the filter function $F(k,\phi)$ introduced in
eq.~(\ref{eq:2.12}) drops to zero on smaller scales, and the further
factor $k$ in the integrand of eq.~(\ref{eq:2.12}) suppresses the
largest scales. Lacking more precise information on the true shape and
normalization of the matter power spectrum on the scales relevant for
the QSO-galaxy cross-correlation function, the normalization to the
observed cluster abundance appears most appropriate for our
purposes. The dependence of $\xi_{\mu\delta}(0)$ especially on the
density parameter is then fairly weak as long as $\Omega_0\ga0.3$. As
Tab.~\ref{tab:1} and Fig.~\ref{fig:4} show, $\xi_{\mu\delta}(0)$
changes by at most a factor of $\la2$ across that range of $\Omega_0$,
and the dependence on $H_0$ is approximately linear. Figure
\ref{fig:7} illustrates the range of QSO-galaxy cross-correlation
functions $\xi_{\rm QG}(\phi)$ that can be covered by varying
cosmological parameters while keeping the normalization of the power
spectrum fixed to the cluster abundance. For illustration, we
overplotted Fig.~\ref{fig:7} with the observational data points by
Ben{\'\i}tez \& Mart{\'\i}nez-Gonz\'alez (1995, 1997).

While non-linear structures on clusters scales are accurately
represented by the non-linear power spectrum, our approach does not
accurately represent the strong lensing that sets in close to cluster
cores. Since we have linearized the magnification factor $\mu$ in
eqs.~(\ref{eq:1.1}) and (\ref{eq:2.1}), our results fail to be
applicable when $\mu$ becomes noticeably larger than unity,
$\mu\ga1.5$, say. This is the case for QSOs closer than $\sim3$
Einstein radii to a cluster core. Depending on cosmological
parameters, QSO and cluster redshifts, and on the velocity dispersion
of the cluster, $\sim3$ Einstein radii correspond to $\sim1-2'$. That
is to say that the expected QSO-galaxy correlation function calculated
here should fall below the true correlation at $\phi\la1-2'$.

The most accurate measurement of $\xi_{\rm QG}(\phi)$ has so far been
obtained by Ben{\'\i}tez \& Mart{\'\i}nez-Gonz\'alez (1995,
1997). They find a significant correlation at angular scales
$\phi<5'$, and a result compatible with zero at larger $\phi$. We can
therefore not yet compare our results to the innermost points of the
observational data. This must be postponed until $\xi_{\rm QG}(\phi)$
has significantly been detected at angles $\ga3'$. We note that
especially the Sloan Digital Sky Survey will provide an ideal data
base for this kind of analysis.

A comparable study was performed by Sanz, Mart{\'\i}nez-Gonzalez, \&
Ben{\'\i}tez (1997) while the work on this paper was completed. In a
forthcoming paper, these authors will perform a detailed comparison of
the data and the theoretical results.

\section*{Acknowledgments}

Special thanks are due to Tsafrir Kolatt, who provided numerical code
for a large variety of power spectra in arbitrary Friedmann
cosmologies. We are further indebted to Enrique
Mart{\'\i}nez-Gonz\'alez, Narciso Ben{\'\i}tez, and Jos\'e-Luis Sanz
for generously sharing their data prior to publication, for numerous
instructive and enjoyable discussions, and for pointing out an error,
and we thank Peter Schneider and Bhuvnesh Jain vor valuable
suggestions. This work was supported in part by the
Sonderforschungsbereich 375 of the Deutsche Forschungsgemeinschaft.

\appendix

\section{The cosmological shear tensor}
\label{app:a}

We summarize in this appendix the derivation of the shear tensor by
which the lensing effects of cosmic material deform thin light
bundles. We start from the conformal Friedmann-Lema{\^\i}tre (FL)
metric,
\begin{equation}
  \d s^2 = a^2(\eta)\,\left[
  \d\eta^2 - \d w^2 - f_K^2(w)\,\d\omega^2\right]\,,
\label{eq:A.1}
\end{equation}
with the scale factor $a$, the conformal time $\eta$, the comoving
distance $w$, and the comoving angular-diameter distance $f_K(w)$,
which depends on the spatial curvature $K$. $\eta$ is related to the
cosmic time $t$ through $\d t=a\d\eta$ with the boundary condition
$\eta=0$ at $t=0$. Measuring lengths in units of the Hubble length
$H_0^{-1}c$, and setting the scale factor at the present epoch to
unity, $K$ can be written in the usual form
\begin{equation}
  K = \Omega_0 + \Omega_\Lambda - 1\;,
\label{eq:A.2}
\end{equation}
where $\Omega_0$ is the matter-density parameter, and $\Omega_\Lambda$
is the density parameter attributed to the cosmological constant, both
taken at the present epoch.

Consider now a fiducial light ray propagating through the space-time
(\ref{eq:A.1}), starting off at the observer into direction
$(\theta_1,\theta_2)$. The affine parameter $\lambda$ along the
fiducial ray is chosen such that $\d\lambda=-a\,\d t$. The transverse
separation vector $\vec\xi$ of a neighboring light ray from the
fiducial ray changes with $\lambda$ according to
\begin{equation}
  \frac{\d^2\vec\xi}{\d\lambda^2} = -\frac{3}{2}\,\Omega_0\,a^{-5}\,
  \vec\xi\;,
\label{eq:A.3}
\end{equation}
(Gunn 1967; Blandford et al.\ 1991; Seitz, Schneider, \& Ehlers
1994). We now substitute in (\ref{eq:A.3}) the comoving separation
vector $\vec x=a^{-1}\vec\xi$ for $\vec\xi$, and the comoving distance
$w$ for $\lambda$. From $\d s=0$ and (\ref{eq:A.1}), $\d w=-\d\eta$
for radial light rays. Since $\d\lambda=-a\,\d t=-a^2\d\eta$, $\d
w=a^{-2}\d\lambda$. After some manipulation, these substitutions yield
\begin{equation}
  \frac{\d^2\vec x}{\d w^2} + K\,\vec x = 0\;.
\label{eq:A.4}
\end{equation}
The solutions of this oscillator equation are the usual linear
combinations of trigonometric (for $K>0$), hyperbolic (for $K<0$), or
linear functions (for $K=0$).

Equation (\ref{eq:A.4}) describes the change in comoving transverse
separation between two light rays on cosmic scales, $\sim
H_0^{-1}c$. Locally, space time is perturbed by density fluctuations
with Newtonian potential $\Phi$. Relative to an unperturbed light ray,
such perturbations deflect light according to
\begin{equation}
  \frac{\d^2\vec x(\vec w)}{\d w^2} = -2\,\nabla_\perp\Phi(\vec w)\;,
\label{eq:A.5}
\end{equation}
(e.g.\ Schneider, Ehlers, \& Falco 1992; Narayan \& Bartelmann 1997)
where we have scaled $\Phi$ by $c^2$ and used that locally, $\d
w=-c\,\d t$. $\vec w$ is the vector
$[f_K(w)\theta_1,f_K(w)\theta_2,w]$. The gradient of $\Phi$ has to be
taken perpendicular to the {\em actual\/} light ray, but for the small
deflections we consider here it can be taken perpendicular to the {\em
unperturbed\/} ray.

Equations (\ref{eq:A.4}) and (\ref{eq:A.5}) describe light deflection
on completely different spatial scales. To combine the effect of the
global curvature of space time with the light deflection by isolated
density fluctuations, we can thus simply add the right-hand side of
(\ref{eq:A.5}) to the right-hand side of (\ref{eq:A.4}). Written in
components, the result is
\begin{equation}
  \frac{\d^2x_i(\vec w)}{\d w^2} + K\,x_i(\vec w) =
  -2\,\Phi,_i(\vec w)\;,
\label{eq:A.6}
\end{equation}
where the index $i$ preceded by a comma denotes the partial derivative
with respect to $x_i$. Equation (\ref{eq:A.6}) is solved by
\begin{equation}
  x_i(\vec w) = f_K(w)\,\vec\theta - 
  2\,\int_0^w\,\d w'\,f_K(w-w')\,\Phi,_i(\vec w')\;,
\label{eq:A.7}
\end{equation}
with the boundary conditions that the unperturbed fiducial ray and the
perturbed ray both start at the observer into directions separated by
$\vec\theta$, hence $\vec x_i(0)=0$ and $\d\vec x_i/\d w=\vec\theta$
at $w=0$. $f_K(w)$ is the comoving angular-diameter distance of the
metric (\ref{eq:A.1}),
\begin{equation}
  f_K(w) = \left\{\begin{array}{ll}
    K^{-1/2}\,\sin(K^{1/2}w) & (K>0) \\
    w & (K=0) \\
    |K|^{-1/2}\,\sinh(|K|^{1/2}w) & (K<0) \\
  		  \end{array}\right.
\label{eq:A.8}
\end{equation}
Recall that (\ref{eq:A.7}) is the comoving separation of a {\em
perturbed\/} light ray from an {\em unperturbed\/} one. The net angle
$\vec\alpha$ by which the perturbed light ray is deflected is
therefore simply
\begin{eqnarray}
  && \alpha_i[f_K(w)\theta_1,f_K(w)\theta_2,w] =
  \frac{x_i(\vec w)-f_K(w)\,\vec\theta}{f_K(w)} \nonumber\\
  &=& -2\,\int_0^w\,\d w'\,\frac{f_K(w-w')}{f_K(w)}\,
  \Phi,_i(\vec w')\;.
\label{eq:A.9}
\end{eqnarray}
(Seljak 1996). Consequently, the shear exerted on a thin light bundle is
\begin{eqnarray}
  \gamma_{ij}(\vec w) &=&
  \frac{\partial\alpha_i[w,f_K(w)\theta_1,f_K(w)\theta_2]}
  {\partial\theta_j} \nonumber\\
  &=& -2\,\int_0^w\,\d w'\,\frac{f_K(w')f_K(w-w')}{f_K(w)}\,
  \Phi,_{ij}(\vec w')\;.\nonumber\\
\label{eq:A.10}
\end{eqnarray}
This result agrees with that of other recent studies which focus on
different lensing statistics (Seljak 1996; Jain \& Seljak 1996;
Bernardeau, van Waerbeke, \& Mellier 1996; Kaiser 1996).


\begin{thebibliography}{99}

\bibitem{B86} Bardeen, J.M., Bond, J.R., Kaiser, N., Szalay, A.S.,
1986, ApJ, 304, 15

\bibitem{BS93a} Bartelmann, M., Schneider, P., 1993a, A\&A, 268, 1

\bibitem{BS93b} Bartelmann, M., Schneider, P., 1993b, A\&A, 271, 421

\bibitem{BS94} Bartelmann, M., Schneider, P., 1994, A\&A, 284, 1

\bibitem{BSH94} Bartelmann, M., Schneider, P., Hasinger, G. 1994,
A\&A, 290, 399

\bibitem{MSB95} Bartelmann, M., 1995, A\&A, 298, 661

\bibitem{BSB96} Bartsch, A., Schneider, P., Bartelmann, M., 1997,
A\&A, 319, 375

\bibitem{BMG95} Ben{\'\i}tez, N., Mart{\'\i}nez-Gonz\'alez, E., 1995,
ApJ, 448, L89

\bibitem{BMG96} Ben{\'\i}tez, N., Mart{\'\i}nez-Gonz\'alez, E., 1997,
ApJ, 477, 27

\bibitem{BW96} Bernardeau, F., van Waerbeke, L., Mellier, Y., 1996,
A\&A, in press; preprint astro-ph/9609122

\bibitem{BS91} Blandford, R.D., Saust, A.B., Brainerd, T.G.,
Villumsen, J.V., 1991, MNRAS, 251, 600

\bibitem{BF93} Bonnet, H., Fort, B., Kneib, J.-P., Mellier, Y.,
Soucail, G., 1993, A\&A, 280, L7

\bibitem{CP92} Carroll, S.M., Press, W.H., Turner, E.L., 1992, ARA\&A,
30, 499

\bibitem{DR87} Dekel, A., Rees, M.J., 1987, Nat, 326, 455

\bibitem{EBW92} Efstathiou, G., Bond, J.R., White, S.D.M., 1992,
MNRAS, 258, P1

\bibitem{FT94} Fischer, P., Tyson, J.A., Bernstein, G.M.,
Guhathakurta, P., 1994, ApJ, 431, L71

\bibitem{FM96} Fort, B., Mellier, Y., Dantel-Fort, M., Bonnet, H.,
Kneib, J.-P., 1996, A\&A, 310, 705

\bibitem{F90} Fugmann, W., 1990, A\&A, 240, 11

\bibitem{G67} Gunn, J., 1967, ApJ, 150, 737

\bibitem{HKLM91} Hamilton, A.J.S., Matthews, A., Kumar, P., Lu, E.,
1991, ApJ, 374, L1

\bibitem{H89} Holtzman, J.A., 1989, ApJS, 71, 1

\bibitem{H95} Hutchings, J.B., 1995, AJ, 109, 928

\bibitem{JS96} Jain, B., Seljak, U., 1996, preprint astro-ph/9611077

\bibitem{JMW96} Jain, B., Mo, H.J., White, S.D.M., 1996, MNRAS, 276,
L25

\bibitem{K84a} Kaiser, N., 1984, ApJ, 284, L9

\bibitem{K92} Kaiser, N., 1992, ApJ, 388, 272

\bibitem{K96} Kaiser, N., 1996, preprint astro-ph/9610120

\bibitem{NB97} Narayan, R., Bartelmann, M., 1997, in: Proc.\ 1995
Jerusalem Winter School, eds. A. Dekel \& J.P. Ostriker. Cambridge:
University Press

\bibitem{MD94} Mellier, Y., Dantel-Fort, M., Fort, B., Bonnet, H.,
1994, A\&A, 289, L15

\bibitem{PD96} Peacock, J.A., Dodds, S.J., 1996, MNRAS, 280, L19

\bibitem{PEI94} Pei, Y.C., 1994, ApJ, 438, 623

\bibitem{PM96} Pell\'o, R., Miralles, J.M., Le Borgne, J.-F., Picat,
J.-P., Soucail, G., Bruzual, G., 1996, A\&A, 314, 73

\bibitem{RH94} Rodrigues-Williams, L.L., Hogan, C.J., 1994, AJ, 107,
451

\bibitem{RH95} Rodrigues-Williams, L.L., Hawkins, M.R.S., 1995, in:
Dark Matter. AIP Conf. Proc. 336, eds. S.S. Holt \& C.L. Bennett (New
York: AIP)

\bibitem{SM97} Sanz, J.L., Mart{\'\i}nez-Gonzalez, E., Ben{\'\i}tez,
N., 1997, preprint

\bibitem{SB95} Schneider, P., Bartelmann, M., 1995, MNRAS, 273, 475

\bibitem{SE92} Schneider, P., Ehlers, J., Falco, E.E., 1992,
Gravitational Lenses. Heidelberg: Springer Verlag

\bibitem{SS95} Seitz, S., Schneider, P., 1995, A\&A, 302, 9

\bibitem{SSE94} Seitz, S., Schneider, P., Ehlers, J., 1994,
Class. Quantum Grav., 11, 2345

\bibitem{UROS} Seljak, U., 1996, ApJ, 463, 1

\bibitem{WF96} Wu, X.-P., Fang, L.-Z., 1996, ApJ, 461, L5

\bibitem{WH95} Wu, X.-P., Han, J., 1995, MNRAS, 272, 705

\end{thebibliography}
\end{document}